\documentstyle{mn}
\input psfig

\ifnfsstwo

\fi

\ifnfssone
  \newmathalphabet{\mathit}
    \addtoversion{normal}{\mathit}{cmr}{m}{it}
    \addtoversion{bold}{\mathit}{cmr}{bx}{it}

\fi

\ifoldfss

\fi

  \makeatletter
  \ifx\CUP@mtlplain@loaded\undefined  
  \makeatother  
    
  \else
    
  \fi

\loadboldmathitalic 
\loadboldgreek      
  
\makeatletter
\ifx\CUP@mtlplain@loaded\undefined  
  \newfont\bit{cmbxti10 at 9pt}
\else
  \newfont\bit{mtbxti10 at 9pt}
\fi
\makeatother

\def\LaTeX{L\kern-.36em\raise.3ex\hbox{a}\kern-.15em
    T\kern-.1667em\lower.7ex\hbox{E}\kern-.125emX}

\newcommand{\gsim}{\mathrel{\hbox{\rlap{\lower.55ex \hbox {$\sim$}}
                   \kern-.3em \raise.4ex \hbox{$>$}}}}
\newcommand{\lsim}{\mathrel{\hbox{\rlap{\lower.55ex \hbox {$\sim$}}
                   \kern-.3em \raise.4ex \hbox{$<$}}}}

\title[Implications of precessing discs and jets]{Observational implications of precessing protostellar discs and jets}

\author[M. R. Bate et al.]
  {M. R. Bate,$^1$\thanks{E-mail: mbate@ast.cam.ac.uk}
   I. A. Bonnell,$^2$
   C. J. Clarke,$^1$
   S. H. Lubow,$^3$
   G. I. Ogilvie,$^{1}$
   \\
   \\\LARGE
   J. E. Pringle$^1$ 
   and C. A. Tout$^1$\\
  $^1$Institute of Astronomy, University of Cambridge, Madingley Road, Cambridge CB3 0HA\\
  $^2$School of Physics and Astronomy, University of St Andrews, North Haugh, St Andrews, Fife, KY16 9SS\\
  $^3$Space Telescope Science Institute, 3700 San Martin Drive, Baltimore, MD 21218, USA
}

\date{Accepted by MNRAS}

\begin{document}

\maketitle

\begin{abstract}

We consider the dynamics of a protostellar disc in a binary system 
where the disc is misaligned with the orbital plane of the binary,
with the aim of determining the observational consequences for such
systems.
The disc wobbles with a period approximately equal to half the
binary's orbital period and precesses on a longer timescale.  We 
determine the characteristic timescale for realignment of the disc 
with the orbital plane due to dissipation.  If the dissipation is 
determined by a simple isotropic 
viscosity then we find, in line with previous studies, that the 
alignment timescale is of order the viscous evolution timescale.  
However, for typical protostellar disc parameters, 
if the disc tilt exceeds the opening angle of the disc, 
then tidally induced shearing within the disc is transonic.
In general, hydrodynamic instabilities associated with the internally
driven shear result in extra dissipation which is 
expected to drastically reduce the alignment timescale.  For large
disc tilts the alignment timescale is then comparable to the precession
timescale, while for smaller tilt angles $\delta$, the alignment timescale
varies as $(\sin \delta)^{-1}$.  We discuss the consequences of the wobbling,
precession and rapid realignment for observations of protostellar jets
and the implications for binary star formation mechanisms.

\end{abstract}


\section {Introduction}

Recent observations of young stellar objects (YSOs) provide evidence 
for the existence of binary systems with circumstellar discs 
that are misaligned with the binary's orbital plane.  The 
Hubble Space Telescope (HST) and adaptive optics images of 
the HK Tau pre-main-sequence system provide the most
striking and direct evidence, assuming that it is indeed a binary system 
(Stapelfeldt et al.\ 1998; Koresko 1998).
In addition, there are several examples of pre-main-sequence binaries
or unresolved YSOs from which two protostellar jets are seen to
emanate in different directions (e.g.\ Davis, Mundt \& Eisl\"offel 1994).  
It is assumed that these twin jets
originate from binary systems with two circumstellar discs that are
misaligned with each other.  Thus, one or both of the discs must be
misaligned with the binary's orbital plane.  Furthermore, 
observations of solar-type main-sequence binary systems indicate that
the stellar rotational equatorial planes in wide binaries ($\gsim 40$ a.u.)
are frequently misaligned with the orbital plane, while for closer
systems there is a tendency for alignment (Hale 1994).  Assuming the
equatorial plane of a star is determined by the plane of its 
original circumstellar disc, this indicates that circumstellar
discs are frequently misaligned with the orbital plane in wide 
binaries.

In binary systems with a circumstellar disc whose plane is misaligned 
with the orbital plane, the circumstellar disc is expected to 
precess due to the tidal interaction of the companion (e.g.\ Papaloizou
\& Terquem 1995).  Disc precession has been used to explain long-period
variations in the light curves of a number of X-ray emitting binary
systems (Gerend \& Boynton 1976; Katz et al.\ 1982;
Wijers \& Pringle 1999).  For YSOs, the major application of disc 
precession has been to predict the precession of protostellar jets,
since many Class 0 and I objects are observed to emit jets 
(Bontemps et al.\ 1996).  Several hydrodynamic investigations into 
the appearance of precessing jets have been performed 
(Raga, Cant\'o \& Biro 1993; Biro, Raga \& Cant\'o 1995; Cliffe et al.\ 1996;
V\"olker et al.\ 1999) and precession has been used to explain
the changes in the flow directions of several jets (Eisl\"offel et al.\ 1996; 
Davis et al.\ 1997; Mundt \& Eisl\"offel 1998), although Eisl\"offel \&
Mundt (1997) stress that alternative explanations exist.  
Recently, Terquem et al.\ (1999) discussed 
the orbital periods that are required to give visibly precessing
jets and the expected precession periods for systems that are
observed to have misaligned jets.

In this paper, we present a simple review of the behaviour of a circumstellar
disc which is misaligned with the orbital plane in a binary system,
and discuss the implications of this behaviour for observations of
discs and jets in binary protostellar systems.
In Section 2, along with simple precession, we also consider
the effect of the oscillating torque produced by the companion,
which could cause the disc and jet to `wobble' with a period of 
half the orbital period.  In Section 3, we consider the effect of
dissipation in the disc and calculate the timescale for a disc 
to realign itself with the binary's orbital plane due to dissipation.
The general results of this study, along with the expected observational
consequences and the implications for models of binary star formation
are discussed in Section 4.  Finally, we give our conclusions in Section
5.  The reader more interested in the results and conclusions of the
paper than the details of the analysis may care to move directly 
to Section 4 from this point.

\section {Precession induced by the binary companion}

We consider here in simple terms the effect of a binary companion (the
secondary) on a gaseous disc around the primary, whose plane is not
aligned with the plane of the binary orbit.

\subsection{ Ring precession}

We first consider the effect of the secondary on a circular ring of material
orbiting the primary at radius $a$.  We consider the binary star system
with component masses $M_p$ (primary) and $M_s$ (secondary), with a
circular orbit, and with stellar separation $D$. Now also consider a
ring of material, of negligible mass, $m_r$, in orbit about the primary
star. The ring has mean radius $a \ll D$, and is tilted to
the plane of the binary orbit at an angle $\delta$. We can regard the
effect of the secondary star on the dynamics of the ring as a
perturbation. We use a set of non-rotating coordinates centred on the
primary star, with the OZ axis parallel to the rotation axis of the
binary. In these coordinates, the acceleration caused by the secondary star at
position vector {\bf r} is given by
\begin{equation}
{\bf F}_s = -(G M_s/D^3) \:  {\bf r} + (3GM_s/D^5) \: ({\bf r.D})
\,{\bf D},
\end{equation}
where ${\bf D}$ is the position vector of the secondary relative to the
primary. We should remark that the fictitious force arising from the
acceleration of the coordinate system has been included, cancelling
the $(G M_s/D^3){\bf D}$ term, and that terms of relative order
$(a/D)$ have been neglected.  Here the first term simply represents an
augmentation of the effect of the gravity of the primary due to the
presence of the secondary. It affects the centrifugal balance of the
ring, but otherwise has no effect on its dynamics. The second term is
the tidal term and represents a force in a direction parallel (or
anti-parallel) to the instantaneous vector joining the two stars, and
with magnitude proportional to the distance of the position {\bf r}
from the plane passing through the primary perpendicular to the line
joining the two stars. From a physical point of view, this term can be
thought of as comprising two components. The first is a force which
acts similarly to a centrifugal force in that it is everywhere
directed perpendicular to and away from the OZ axis, and in that its
magnitude is proportional to the distance from that axis. In terms of
azimuthal Fourier components $\exp(i m \phi)$, where $\phi$ is the
azimuth about the OZ axis, this term is derived from an axially
symmetric potential, and thus has $m=0$. The second is similar except
that the magnitude of the force is proportional to distance from the
OZ axis multiplied by $\cos(2 \phi)$, and thus has $m=2$.

\subsubsection{The $m=0$ term}

Breaking the tidal term into these two parts enables us to consider
their effects on the dynamics of the ring in a more straightforward
manner. Suppose that the coordinates of the secondary are given by 
\begin{equation}
{\bf D} = ( - D \sin(\Omega_b t),  D \cos(\Omega_b t), 0),
\end{equation}
where $2 \pi /\Omega_b$ is the binary period, and thus
\begin{equation}
\Omega_b = (G (M_s+M_p)/D^3)^{1/2}.
\end{equation}

Suppose that the ring is tilted in our
system of coordinates about the OX axis. Thus the axis of the ring is
given by
\begin{equation}
{\bf k} = (0, - \sin\delta, \cos\delta).
\end{equation}
Then the effect of the $m=0$ part of the force is to exert a torque
${\bf G}_0 = (G_{0x},0,0)$ on the ring about the negative OX axis, of
magnitude
\begin{equation}
G_{0x} = - (3/4) \, ( G M_s m_r a^2 /D^3) \,  \sin\delta \cos\delta,
\end{equation}
where $m_r$ is the mass of the ring. The direction of the $m=0$
torque, $\bf {G}_0$, lies along the line of nodes of the ring (the
line along which the ring intersects the OXY-plane). If ${\bf \Omega}
= \Omega {\bf k}$ is the angular velocity of the ring, then ${\bf G}_0 
\cdot {\bf \Omega} = 0$. Thus the effect of the torque on the
ring is to cause the plane of the ring to precess retrogradely about
the OZ axis with a precession rate given by $-\omega_p$, where
\begin{equation}
\omega_p = \mid G_{0x} \mid /(\sin \delta \, m_r a^2 \Omega).
\end{equation}
Note that  $\Omega$ is  given by
\begin{equation}
\Omega = {(G M_p /a^3)}^{1/2},
\end{equation}
and $m_r a^2 \Omega$ represents the angular momentum of the
ring.

Thus we find that the mean precession rate is 
\begin{equation}
\omega_p = (3/4) \, \cos\delta \, (G M_s /D^3  \Omega),
\end{equation}
or, equivalently, that
\begin{equation}
\omega_p/\Omega = (3/4) \, \cos\delta \, (M_s/M_p) \, (a/D)^3.
\end{equation}
We should also note that the above treatment, which essentially
assumes that the ring maintains its shape as if it were a solid body, 
will be valid
provided that $\Omega \gg \Omega_b \gg \omega_p$, which is satisfied in
this case (see below).

\subsubsection{The $m=2$ term}
\label{m2term}

The effect of the $m=2$ term is to give rise to an oscillating torque
\begin{equation}
{\bf G}_2 = (G_{2x}, G_{2y}, G_{2z}) 
\end{equation}
on the ring which has a frequency of $2 \Omega_b$.
The $m=2$ torque is of the form
\begin{equation}
G_{2x} = - (3/4) \, (G M_s m_r a^2 /D^3) \,  \sin\delta \cos\delta \cos(2 \Omega_b t),
\end{equation}
\begin{equation}
G_{2y} =  - (3/4) \, (G M_s m_r a^2 /D^3) \,  \sin\delta \cos\delta \sin(2 \Omega_b t),
\end{equation}
and
\begin{equation}
G_{2z} =  - (3/4) \, (G M_s m_r a^2 /D^3) \,  \sin^2\delta \sin(2 \Omega_b t).
\end{equation}
We note also that ${\bf G}_2 \cdot {\bf \Omega} = 0$. Indeed, the
total torque on the ring may be expressed simply as
\begin{equation}
{\bf G} = (3/2) (G M_s m_r a^2/D^5) ({\bf D} \cdot {\bf k}) ({\bf D}\times{\bf k}).
\end{equation}
Because of the
oscillating nature of the $m=2$ torque, the mean precession rate is
unaffected by the $m=2$ term.  However, the instantaneous precession
rate oscillates with a period of (approximately) one half of the
orbital period (Katz et al.~1982). Since the amplitude of the torque
due to the $m=2$ term is equal to the torque generated by the $m=0$
term the modulation is not a negligible effect.

Perhaps the simplest way to envisage what is happening is to consider
the motion of the symmetry axis ${\bf k}$ of the ring. The $m=0$ term
causes the axis to move in a cone, semi-angle $\delta$, around the OZ
axis at a rate $ - \omega_p$. Superimposed on this, the $m=2$ term
causes a wobble of the symmetry axis in the direction of precession,
and a nodding motion perpendicular to this direction. The amplitude of
the wobble (in radians) is roughly $ \omega_p/(2 \Omega_b)$, and the
amplitude of the nodding motion is the same but multiplied by a factor
of $\tan\delta$ (see Katz et al.~1982). 
We should remark that this does not lead to a
divergence as $\delta \rightarrow \pi/2$ because $\omega_p$ is
proportional to $\cos\delta$. We should note that, strictly, because
the ring is precessing in a retrograde direction because of the $m=0$
term, the period of the wobble is in fact $2 \pi/(2 \Omega_b +
\omega_p)$.

\subsection{Disc precession}

We have seen above that the precession rate for a ring of radius $a$
is proportional to $a^{3/2}$. Thus if we regard a disc as being made
up of a succession of concentric rings, the effect of the $m=0$
potential is to cause the disc to precess differentially. However, if
the disc is able to hold itself together in some way, either by means
of wave-like communication (Larwood et al.~1996), or by viscous
communication (Wijers \& Pringle 1999), then it may be able to respond
coherently to the $m=0$ term in the potential by precessing at the
same rate at all radii, that is, as if it were a solid body. The
precession rate can then be calculated straightforwardly from the
above, by rewriting the ring mass, $m_r$, as the mass of an annulus in
the disc, $2 \pi \, \Sigma \, a da$, where $\Sigma$ is the disc
surface density at radius $a$.

The net precession rate is then given by
\begin{equation}
\omega_p = K \cos\delta \, (G M_s/D^3 \Omega_d),
\end{equation}
where (Terquem 1998)
\begin{equation}
K = \frac{3}{4} R^{-3/2} \, \left[\int_{0}^{R} \Sigma a^3 da\right] / \left[\int_{0}^{R} \Sigma a^{3/2} da\right].
\end{equation}
Here $\Omega_d$ is the angular velocity of disc material at the outer
edge of the disc which is at radius $R$, and the lower limit in the
integrals can be simply a small radius rather than zero.

Thus, for a disc,
\begin{equation}
\label{precessionrate}
\omega_p/\Omega_d = K \cos\delta \, (M_s/M_p) (R/D)^3.
\end{equation}
For a constant surface density disc $K=15/32$ (Larwood et al.\ 1996); 
for a surface density proportional to $R^{-3/2}$, $K=3/10$.

\subsection{ Application to protostellar discs}

In protostellar discs the ratio of disc semi-thickness to radius,
$H/R$, is of order 0.1 (Burrows et al.~1996; Stapelfeldt et al.~1998), 
which is larger than the typical dimensionless viscosity 
$\alpha \sim 0.01 $ (Hartmann et al.~1998).  Under these
circumstances bending waves can propagate through the disc (assumed
Keplerian) (Papaloizou \& Lin 1995, Pringle 1997), and thus the disc
can communicate with itself in a wave-like manner.  The propagation
speed for such bending waves is of order $\frac{1}{2} c_s$, where
$c_s$ is the disc sound speed (Papaloizou \& Lin 1995, Pringle
1999). Thus the condition that the disc be able to precess as a solid
body subject to forced differential precession might be expected to
be that the wave crossing timescale be less than the precession
timescale, that is
\begin{equation}
R /c_s \lsim \omega_p^{-1}.
\end{equation}
This is the criterion given by Larwood et al.\ (1996), and by Papaloizou
\& Terquem (1995). Using the standard result for accretion discs
(e.g.~Pringle 1981) that $c_s/(R \Omega_d) \sim H/R$, we find that the
condition may be written
\begin{equation}
\omega_p/\Omega_d \lsim H/R.  \label{condition1}
\end{equation}

For a disc in a binary system which is truncated by tidal forces, the
outside disc edge is typically a substantial fraction of the binary separation
(e.g Papaloizou \& Pringle 1977; Paczy\'nski 1977; Artymowicz \& Lubow 1994; 
Larwood et al.\ 1996). 
As a typical value, for
reasonable mass ratios, we shall take $R/D \approx 0.3$. Thus, taking
$K \approx 0.4, q = M_s/M_p \approx 1$, and $\cos\delta \approx 1$,
we find that
\begin{equation}
\omega_p/\Omega_d \approx 0.011\; (K/0.4) (R/0.3D)^3 \, q \, \cos\delta,
\end{equation}
We note that condition (\ref{condition1}) is readily satisfied for
protostellar discs. However, as we shall note below, this is not the
whole story.

Using the above estimates, we note that
\begin{equation}
\begin{array}{ll}
\hspace{-5pt} \Omega_d/\Omega_b\hspace{-8pt}& = [M_p/(M_p+M_s)]^{1/2} (D/R)^{3/2} \\
\\
& \approx 4.3\; [2/(1+q)]^{1/2}\; (R/0.3 D)^{-3/2}.
\end{array}
\end{equation}
Thus typically
\begin{equation}
\label{omega_p_Omega_b}
\omega_p/\Omega_b \approx 0.05\; (K/0.4)\; q[2/(1+q)]^{1/2}(R/0.3 D)^{3/2}\cos\delta.
\end{equation}
This implies that the amplitude of the wobble/nodding motion 
about steady precession (discussed in Section \ref{m2term}) is of order 
a degree or so at the outer edge of
the disc although, as we note in Section \ref{precessjet}, 
the amplitude inside the disc depends on the properties of the disc.

Some protostellar discs may be sufficiently massive that a noticeable
deviation from Keplerian rotation may occur.  This effect, and also the
self-gravitation of the disc, could change the nature of the propagating
bending waves, making them dispersive (Papaloizou \& Lin 1995).  However,
assuming that self-gravity is not dominant (i.e.\ the Toomre 
parameter $Q \ga 1$), these effects are expected
to be small because the fractional deviation from Keplerian rotation is at
most of order $H/R$, and our estimates here are probably still valid.
The character of wave propagation also differs between vertically isothermal
and thermally stratified discs \cite{LubOgi1998}, however this is 
relevant only when the radial wavelength is comparable to the disc 
thickness, and therefore does not affect the propagation of 
long-wavelength bending waves as considered here.

\section{The effect of dissipation}

As we have seen, the tidal effect of the secondary star can be
regarded being due to the $m=0$ and $m=2$ parts of the tidal potential
independently. Thus the effect of dissipation within the disc on the
motions caused by these two parts of the potential can also be
calculated separately.

\subsection{The $m=0$ term}

The effect of the $m=0$ term on the misaligned disc is to produce a
torque ${\bf G}_0$ on the disc which is in the direction of aligning it
with the binary orbit. This comes about because there is a potential
energy, $\Phi$, associated with the misalignment of the disc which is
given by
\begin{equation}
\Phi = (3 G M_s/ 8 D^3) \sin^2 \delta \, \int_{0}^{R} a^2 \Sigma 2 \pi a da.
\end{equation}
This has minima at $\delta=0$ (alignment) and $\delta= \pi$
(anti-alignment), and a maximum at $\delta = \pi/2$.  Thus we would
normally expect loss of energy associated with the motions induced by
the $m=0$ part of the tidal potential to lead to the disc becoming
aligned with the orbital plane.

However, because the $m=0$ term is symmetric about the OZ-axis, ${\bf
\hat{z}}$, any torque produced by the $m=0$ term must lie in the
OXY-plane. We have seen that in the absence of dissipation the torque
is such that ${\bf G}_0$ is parallel to ${\bf k \times \hat{z}}$,
where ${\bf k}$ is the unit vector in the direction of the angular
momentum ${\bf J}$ of the disc. The effect of viscosity is to give
rise to a phase delay in the torque about the OZ axis, which has the
effect of producing an additional (viscous) torque ${\bf G}_{0 \nu}$,
which is perpendicular to $\bf \hat{z}$, which lies in the plane
defined by ${\bf J}$ and ${\bf \hat{z}}$, and which is directed such
that ${\bf G}_{0\nu} \cdot {\bf J} < 0$, so that its effect is
dissipative. Thus the net effect of this torque is to align the disc
with the OZ-axis, but in such a way that ${\bf J \cdot \hat{z}}$ is
conserved. This means that in the process of alignment, angular
momentum is removed from the disc, giving rise to an enhanced
accretion rate (see Papaloizou \& Terquem 1995). We compute the
timescale on which this torque leads to disc alignment below.

\subsection{The $m=2$ term}

In the absence of dissipation, the effect of the $m=2$ term is to
induce oscillations in the disc which have a frequency of $2
\Omega_b$, and which therefore cause the angular momentum vector ${\bf
J}$ to oscillate about some mean value. Thus we would naively expect
that the effect of dissipation on the motions induced by the $m=2$
part of the tidal potential would be to reduce the amplitude of the
oscillations, but to have no net effect on the mean disc plane, that
is, to have no long-term time-averaged effect on the value of
$\delta$.

Papaloizou \& Terquem (1995) and Terquem (1998) have argued that the
effect of dissipation on the $m=2$ induced motions is to give rise to a
torque which may {\it increase} the inclination of the disc.  This
possibility has been investigated in detail by Lubow \& Ogilvie 
\shortcite{LubOgi2000},
who formulated the problem in terms of the linear stability of an
initially coplanar disc in the presence of the tidal field.  It was
confirmed that the $m=2$ component of the tidal potential causes the
inclination to grow, but the effect is usually negligible and is
outweighed by the effect of the $m=0$ potential, so the net outcome is
that the inclination decays in time.  An exception occurs if there is a
coincidence between the frequency of the oscillating torque, $2\Omega_{\rm b}$,
and the natural frequency of a global bending mode of the disc.  However,
this resonance can occur only if the disc is very thick ($H/R\ga0.4$) or
much smaller than the standard tidal truncation radius ($R/D\ll0.3$).  We
therefore neglect the effect of the $m=2$ term on the long-term 
time-averaged value of $\delta$ and restrict our
attention to the alignment effects of the $m=0$ term.

\subsection{Application to protostellar discs}
 
In the analysis so far we have assumed that the disc is able to
communicate internally sufficiently fast that the disc can precess as
a rigid body.  This communication takes the form of bending waves
which propagate through the disc at a speed of order
${\textstyle{{1}\over{2}}}c_{\rm s}$.  However, these bending waves
are associated with nearly resonant horizontal epicyclic motions that
are strongly shearing, being proportional to the distance above the
mid-plane (Papaloizou \& Pringle 1983).  Viscosity in the disc can
then act on these shearing motions, leading to dissipation of energy
in the precessional motion, and alignment of the disc with the orbital
plane.

We have seen that the torques exerted at different radii in the disc
by the $m=0$ component of the tidal potential would naturally result
in differential precession.  In order for the disc to resist this,
hydrodynamic stresses must be established within the disc so that the
net torque on each ring is such as to maintain a uniform, global
precession rate.  The required internal torque, although it varies with
radius and vanishes at the edges of the disc, is therefore generally
comparable to the total tidal torque on the disc, and is given
approximately by
\begin{equation}
G_{\rm int}\sim2\pi\Sigma R^4\Omega_{\rm d}\omega_{\rm p}\sin\delta.
\end{equation}
These hydrodynamic stresses are associated with the horizontal
epicyclic motions mentioned above, which take the form
\begin{equation}
v_r^\prime\sim 2v_\phi^\prime\sim Az,
\end{equation}
referred to cylindrical polar coordinates $(r,\phi,z)$ based on the
mean disc plane, and where $A$ is independent of $z$.  As a result of
the hydrodynamic stress $\rho v_r^\prime\cdot r\Omega$ there is a net
horizontal angular momentum flux
\begin{equation}
2\pi r\int\rho v_r^\prime r\Omega z\, dz\sim2\pi A\Sigma
R^2H^2\Omega_{\rm d}.
\end{equation}
Equating this with $G_{\rm int}$ gives
\begin{equation}
A\sim(R/H)^2\omega_{\rm p}\sin\delta,
\end{equation}
and so
\begin{equation}
v_r^\prime/c_{\rm s}\sim2v_\phi^\prime/c_{\rm s}\sim(R/H)^2
(\omega_{\rm p}/\Omega_{\rm d})\sin\delta(z/H).
\end{equation}
Note that for these velocities to be subsonic we require that
\begin{equation}
\label{openingangle}
\omega_p/\Omega_d \lsim  (H/R)^2 / \sin\delta.
\end{equation}
For $\sin\delta > H/R$, this is a stronger condition than the one
derived by Papaloizou \& Terquem (1995), and if $\delta \sim 1$, it is
only marginally satisfied in protostellar discs. The fact that the
internal disc velocities are likely to be close to sonic, means that
the dissipation may well be strongly enhanced. We discuss this further
below.

For an internal disc viscosity $\nu$, these velocity perturbations
lead to a rate of energy dissipation in the disc of order
\begin{equation}
dE/dt \sim m_d c_s^2 \,(\nu/H^2) \, (R/H)^4 (\omega_p/\Omega_d)^2 \sin^2\delta,
\end{equation}
where $m_d$ is the mass of the disc.

As we have seen above, the effect of this dissipation is to lead to an
alignment of the disc with the orbital plane. The means by which this
is accomplished is through the viscous torque ${\bf G}_{0\nu}$, which
leads to alignment by removing the component of disc rotation which
lies in the orbital plane (recall that ${\bf G}_{0\nu} \cdot {\bf
\hat{z}} = 0$). Thus the amount of energy to be dissipated in order to
bring about alignment is
\begin{equation}
E_{\rm kin} \sim m_d R^2 {\Omega_d}^2 \sin^2 \delta. 
\end{equation}

From these estimates we deduce the alignment timescale for the disc
${\rm t_{align}}$, which is given by
\begin{equation}
{\rm t_{align}} \sim E_{\rm kin}/(dE/dt) \sim (R^2/\nu) \,  (H/R)^4
(\Omega_d/\omega_p)^2 .
\end{equation}
We note further that the viscous evolution timescale for a disc $t_\nu$
is given by
\begin{equation}
t_\nu \sim  R^2/\nu.
\end{equation}
Thus, {\it from this analysis} we would conclude that for precessing discs
in which the induced velocities are subsonic, the alignment timescale
for an misaligned protostellar disc is typically of order, or somewhat
longer than, the viscous evolution timescale, and that the additional
accretion rate caused by the process of alignment is typically at most
comparable to the accretion rate already present in the disc.

The viscous evolution timescale can be written in terms of the
dimensionless measure of viscosity, $\alpha$, as
\begin{equation}
t_\nu \sim {\Omega_d}^{-1} (R/H)^2 \alpha^{-1}.
\end{equation}
Using this we find that 
\begin{equation}
t_{\rm align} \sim {\omega_p}^{-1}  \alpha^{-1} (H/R)^2 (\Omega_d/\omega_p).
\end{equation}
This expression for the alignment timescale and the expression for
the precession rate (equation \ref{omega_p_Omega_b}) have been 
verified by the numerical calculations of Lubow \& Ogilvie 
\shortcite{LubOgi2000}.

\subsubsection{The effect of sonic induced velocities}

However, since the velocities associated with the induced shearing
in these discs are close to sonic, the parametric 
instabilities discussed by Gammie, Goodman \& Ogilvie (2000) are likely
occur.  These authors concluded that
the internal shear motions induced by the disc bending are, for sonic
shearing motions, unstable on a local (disc) dynamical timescale,
$\Omega^{-1}$. The local turbulence induced by these instabilities
increases the damping rate for the shearing motions. 
For example, if $v' \sim c_s$, then we may estimate the
dissipation rate in the disc due to these instabilities as
\begin{equation}
dE/dt \sim m_d   {c_s}^2 \, {\Omega_d},
\end{equation}
which implies an alignment timescale of
\begin{equation}
t_{\rm align} \sim {\omega_p}^{-1} / \sin \delta.
\end{equation}
This implies that alignment occurs almost as fast as precession.

Furthermore, even when the induced velocities, $v'$ are subsonic, the
instabilities grow on a timescale $\sim H/v'$
(Gammie et al.\ 2000), provided that the growth time is less than the 
viscous damping timescale $\sim H^2/\nu$, i.e. provided that
\begin{equation}
\alpha \lsim v'/c_s,
\end{equation}
or, alternatively, that
\begin{equation}
\sin \delta  \gsim (\Omega_d/\omega_p) \, (H/R)^2 \: \alpha
\end{equation}
i.e.\ that
\begin{equation}
\sin \delta \gsim 0.01 \frac{\mbox{$(\alpha/0.01) \, (H/0.1R)^2 $}}{\mbox{$(K/0.4) \, (R/0.3D)^{3} \, q \, \cos \delta$}}.
\end{equation}
In this case the parametric instabilities again give
enhanced dissipation which is likely to lead to an enhanced effective
viscosity, $\alpha_{\rm eff}$, such that the enhanced damping rate is
of order the instability growth rate, {\it viz.}
\begin{equation}
\label{efficiency}
\alpha_{\rm eff} \sim v'/c_s.
\end{equation}
This enhanced damping implies an alignment timescale of 

\begin{equation}
\begin{array}{ll}
\hspace{-5pt} t_{\rm align} \hspace{-6pt} & \sim {\omega_p}^{-1} (H/R)^4 \, (\Omega_d/ \omega_p)^2 / \sin \delta \\
\\
& \sim {\omega_p}^{-1} \frac{\mbox{$(H/0.1R)^4 \, (K/0.4)^{-2} \, (R/0.3D)^{-6} \, q^{-2} $}}
{\mbox{$\cos^2 \delta \sin \delta$}}.
\end{array}
\end{equation}

We should stress that this estimate is probably a lower limit to the
alignment timescale since the efficiency of the damping due to the
parametric instability may not be as perfect as suggested by equation
\ref{efficiency} (Gammie et al.\ 2000).
We also note that although Gammie et al.\ (2000) considered an 
isothermal, Keplerian disc,
the growth rate of the parametric instability is not sensitive to these
assumptions.  The instability would also act in a thermally
stratified disc, or in a self-gravitating disc with slightly non-Keplerian
rotation.

\section{Discussion}

\subsection{General results}

We have presented a simple discussion of the dynamics of a misaligned
disc in a binary star system, and have applied the results to the
parameters of discs and binary parameters which are relevant to young
stellar objects. Our general conclusions are as follows:

For typical protostellar disc parameters, and for circular binary orbits, tidal
forces cause a misaligned disc to precess like a solid body about the
angular momentum vector of the binary orbit, with a precession period
of order $P_p \sim 20 P_{orb}$, where $P_{orb}$ is the orbital period
(equation \ref{omega_p_Omega_b}; Papaloizou \& Terquem 1995; 
Larwood et al.~1996; Terquem 1998; 
Terquem et al.~1999). In addition, the outer disc plane is forced
to wobble with a period of $\frac{1}{2} P_{orb}$ (Katz et
al.~1982).

We have presented order of magnitude estimates for the timescale on
which dissipation within the disc leads to alignment with the orbital
plane. (For the reasons described above, we overlook 
the possibility raised by Papaloizou \& Terquem (1995) that
dissipation might lead to disc misalignment). If the disc evolution
is determined by a simple isotropic viscosity, then we find, in line
with the estimates given by Papaloizou \& Terquem (1995) and by
Terquem et al.\ (1999) that the alignment timescale is, for protostellar
disc parameters, of order the normal viscous evolution
timescale. 

However, we have also pointed out that the velocities induced within
the disc by the action of tidal forces on the tilt are, for typical
protostellar disc parameters, transonic, and that the criterion used
by Papaloizou \& Terquem (1995) to justify the use of their
linearization procedure is incorrect if the disc tilt exceeds the
opening angle of the disc (equation \ref{openingangle}). 
The induced velocities take
the form of a horizontal epicyclic motion proportional to the distance
above the mid-plane within the disc which oscillates in a frame rotating
with the fluid
with a period approximately equal to that of the orbital period
of the disc material. It has long been suspected that such a shear
flow is unstable (Kumar \& Coleman 1993), and recent work by 
Gammie et al.\ (2000) has demonstrated that the flow is indeed 
unstable to a parametric hydrodynamic instability which has a 
growth rate of order the shearing timescale and leads to rapid dissipation. 
We have estimated the effect of such instabilities. For large
disc tilts, and for typical protostellar disc parameters, we find
that the disc alignment timescale is comparable to the precession
timescale. However, for smaller tilt angles $\delta$, we find that the
alignment timescale varies as $(\sin \delta)^{-1}$.  It is worth
noting that the enhanced dissipation will also result in a greater disc
luminosity and a larger mass accretion rate than those provided by standard
viscous evolution.  These effects should be considered when modelling 
protostellar discs in binary systems.  However, given the large range
of observed accretion rates from protostellar discs
(e.g.~Hartmann et al.~1998) and the uncertainties in current models 
of protostellar discs, it would be difficult to detect 
an unambiguous signature of the enhanced luminosity or accretion rate.

We should draw attention to the fact that, in common with previous
authors, all the results presented above are for binary stars with
circular orbits. If, as is the case for many binary stars, the orbit
is non-circular, then the analysis is similar but rather more
complicated. However, the main effects of orbital eccentricity on the
results reported above are to decrease the ratio of disc radius to
orbital semi-major axis and to modify the time-averaged potential
due to the companion.  Tidal truncation of the disc now takes
place at periastron, so that $R \approx 0.3 a (1-e)$, while the
modified time-averaged potential can be
approximated by replacing the orbital separation $D$, by 
$a(1-e^2)^{1/2}$ (Holman, Touma \& Tremaine 1997), 
where $a$ is the semi-major axis and $e$ 
is the eccentricity.  
Thus, to a first approximation, the major effects of orbital 
eccentricity can be taken into account by replacing $R/0.3D$ in 
the above formulae by the quantity
\begin{equation}
\sqrt{\frac{1-e}{1+e}}\left(\frac{R}{0.3a(1-e)}\right).
\end{equation}

\subsection{Precession and wobbling of protostellar jets}
\label{precessjet}

Disc precession has been used as an ingredient in the explanation of
long-period variations in the light curves of a number of X-ray
emitting binary systems (Gerend \& Boynton 1976; Katz et al.\ 1982;
Wijers \& Pringle 1999). However, for protostellar systems the major
application for tidally induced disc precession has been to provide
an explanation for changes in flow direction of protostellar jets
(e.g.\ Eisl\"offel \& Mundt 1997; Terquem et al.\ 1999). Although changes in
flow direction are often discussed in terms of precession, there are,
as discussed by Eisl\"offel \& Mundt (1997), various alternative
explanations for such phenomena, and there is as yet no convincing
case of a jet which has been steadily precessing for many precession
periods. In the light of this we discuss, in general terms, the kind
of effects which tidally induced disc precession might be expected to
lead to from a theoretical point of view.

Protostellar jets appear to be produced during the major accretion
phase in the star-formation process, that is during the Class 0 and
Class I phases of the life of a protostellar core/young star (Bontemps
et al.\ 1996). This
phase of the star formation process is thought to last about 
$10^5$ years (Lada 1999). 
In addition, given that the jet velocities are of order
the escape velocities from the central stars, and that the mass
outflow rates are a non-negligible fraction of the likely accretion
rates, it seems a reasonable assumption that the jets are formed close
to the centre of the disc (e.g.\ Pringle 1993), and therefore that the
jet direction is governed by the disc axis in the central regions of
the disc. Thus, since for a strongly misaligned disc, the alignment
timescale is of order the precession timescale, which is of order
$\sim 20$ orbital periods, we expect that, regardless of how the
misalignment might have come about, strongly misaligned discs and jets are only
likely to occur in binaries with orbital periods longer than $\sim 5000$
years, that is with separations larger than about $\sim 100$
a.u.  Examples of systems which appear to have misaligned jets include:
Cep E \cite{Eisloffel96}; T Tau \cite{BohSol94}; HH 1,2,144 VLA 1/2
\cite{Reipurth93}; and HH 111/121 \cite{GreRei93}.  These systems are
either known to be wide binaries ($\gsim 100$ a.u.), in agreement
with expectations, or have not yet been resolved.

Since the alignment and precession timescales are comparable for strongly 
misaligned discs, one would not expect to
observe the multiple large scale wiggles which might be the result of
a jet undergoing many precession cycles at a large angle to the
orbital axis. One might, however, see the results of such a jet whose
direction starts at large angle to the binary axis, and then changes
direction on the sky as it simultaneously precesses and aligns (with
the binary axis) on a timescale of $\sim 20$ orbital periods. For such
a jet, with jet axis at a large angle to the binary axis, provided that
the inner and outer parts of the disc are in good communication, the
jet direction is forced to wobble by the $m=2$ tidal component with a
period of half the orbital period. The amplitude of the
wobble at the outside of the disc is only about a degree or so, but
since the communication to the disc centre is wave-like, the amplitude
of the wobble at the disc centre must depend on the wave amplitude
induced there, which in turn depends on the properties of the disc,
and in particular on the dependences with radius of the angular
momentum density ($\propto \Sigma r^{1/2}$) and the group velocity
($\propto c_s$) (e.g.\ Terquem 1998). Such a regular wobbling of the
jet direction may already have been observed in one
of the jets emanating from Cep E \cite{Eisloffel96}.  The jet appears to 
undergo a regular wobble with an amplitude of $4^{\circ}$ and a 
period of $\sim 400$ years (both dependent on the inclination).  
The mechanism driving this oscillation 
may be able to be tested observationally using the VLA or adaptive optics
in the infrared.  If the oscillation of the jet
is due to a wobble of the disc, the period of the binary should be
$\sim 800$ years (separation $\sim 80$ a.u. or $0.12''$), whereas if
the oscillation is due to precession of the disc the binary's period 
should be much shorter ($\lsim 20$ years) and the binary would not be 
resolvable ($\lsim 0.01''$).  Many YSOs with single jets
also show evidence for direction changes, and a high-resolution survey
to determine the binarity of such sources would be invaluable for 
testing the theory of wobbling and precessing jets.

For jets that are weakly misaligned, we note that the alignment
timescale is inversely proportional to $\delta$, (implying that the
misalignment angle decreases with time 
as $t^{-1}$) and scales, for typical protostellar
disc parameters, approximately with the orbital period.  Furthermore, the
ratio of the precession timescale to the alignment timescale scales
with $\delta$.  In order to see multiple wiggles caused by jet
precession in a jet whose length corresponds typically to a dynamical
timescale of $\sim 10^4$ years, we require a precession period of
less than or of order a few thousand years, and thus binary periods of
less than or of order a few hundred years, and binary separations less
than or of order a few tens of a.u.  For example, an initially 
strongly misaligned binary with a separation of a few tens of a.u. will
rapidly evolve into a weakly misaligned system, but even after 
$\sim 10^5$ years will still show a misalignment angle of $\sim 0.03$ radians, 
that is, a few degrees, and thus, with a precession period of a few 
thousand years, such a system would be marginally observable.  
Wider binaries than this would have precession periods too
long to show wiggling of a jet whose dynamical age is $\sim 10^4$
years, whilst closer binaries with smaller precession periods would
have alignment angles too small to be observed. 
We note that for small
misalignment angles, the amplitude of the wobble of the outer disc
caused by the $m=2$ tidal term (which is approximately 
$\delta \omega_{\rm p}/(2\Omega_{\rm b}) \approx \delta/40$ as
$\delta \rightarrow 0$) is probably too small to be detected.

In summary, one expects jets from strongly misaligned discs to be a
rarity amongst binaries closer than $\sim 100$ a.u.  In wider binaries,
jets in strongly misaligned systems are wiggled at twice the binary
orbital frequency with an amplitude which depends on the details of
the disc structure.  Weakly misaligned systems are longer-lived, and
therefore potentially observable in closer binaries.  As $\delta
\rightarrow 0$, precession should give rise to observable jet wiggling
(with amplitude of angle $\delta$ decreasing as $\propto
\delta^{-1}$) in binaries with separation of order a few tens of a.u.

\subsection{Implications for binary star formation}

We now consider briefly the implications that evidence for disc
misalignment might have for the various theories of binary star formation,
noting that in order for a misaligned system to be observed, the system as a 
whole must be assembled over a timescale which is shorter than 
the alignment timescale, $t_{\rm align}$.

\subsubsection{Binary fragmentation}

Most published fragmentation calculations result in discs which are 
coplanar with the resulting fragments.
Of these, there are typically two cases: fragmentation due to initial
density perturbations (e.g.~Boss \& Bodenheimer 1979; Boss 1986), or
centrifugally-supported fragmentation (e.g.~fragmentation of a 
massive protostellar ring or disc; Norman \& Wilson 1978; Bonnell 1994;
Bonnell \& Bate 1994; 
Burkert \& Bodenheimer 1996; Burkert, Bate \& Bodenheimer 1997) with some
calculations exhibiting both types of fragmentation 
(e.g.~Bonnell et al.~1991; Bate, Bonnell \& Price 1995).  The
fragmentation of centrifugally-supported material leads to the orbit(s)
and discs of all fragments occupying the same plane.  
In most calculations where
the fragmentation occurs due to initial density perturbations,
the initial conditions have a single axis of 
rotation and there is typically an $m=2$ density perturbation which 
is perpendicular to the rotation axis.  With these simple initial 
conditions, a binary forms from the initial $m=2$ density 
perturbation in a plane perpendicular to the rotation axis and passing
through the centre of the cloud, while the discs that form around 
these fragments
have rotation axes that are parallel to the rotation axis of the cloud.
Thus, the protostellar discs lie in the same plane as the orbit.

In order to produce discs that are misaligned with the orbital plane, 
it is natural to expect that the angular momentum distribution of the
molecular gas must have strong spatial variations to enable the discs to
have different rotation axes from the larger-scale orbit.  However, this is
not necessary.  All that is required is to force the binary to form
in a plane that is not perpendicular to the rotation axis.  This can 
be achieved simply by rotating the $m=2$ density perturbation slightly so that
it is no longer perpendicular to the rotation axis (or visa versa).
Such a lack of correlation between the axis of rotation and the initial
density perturbation(s) is expected if the collapse of a molecular
gas is triggered by an external source (e.g.\ a shock wave or gravitational
interaction with a passing object), or if the pre-collapse clump of gas
is formed dynamically within a turbulent molecular cloud.  In fact,
given that the rotational energy of observed molecular clumps
is generally insignificant compared with their gravitational energy
(Goodman et al.\ 1993), it is hard to find a reason why there 
should be any correlation.
In this case, the two fragments form above and below the plane that 
is perpendicular to the rotation axis and passes through the cloud's 
centre and, hence, the orbital plane is no longer perpendicular to the rotation
axis of the initial cloud.  The rotation axes of the discs that 
form around the fragments, on the other hand, are still parallel 
to the rotation axis 
of the initial cloud.  Thus, even though the pre-collapse gas is all
rotating around a single axis, possibly in solid-body rotation, 
the discs are misaligned with the orbital plane of the binary.
Such fragmentation calculations have been performed by Bonnell et al.~(1992)
and Bonnell \& Bastien 
(1992).  In their particular case, the $m=2$ density perturbation 
was in the form of the initial molecular cloud core being prolate.
We note that in such calculations, although the discs are misaligned with
the orbital plane, they are still aligned with each other.  To produce
the added complexity of discs which are {\it initially} misaligned with 
each other would
require spatial variations in the angular momentum distribution.  However,
even if the discs are initially aligned, they are almost certain to 
undergo precession at different rates (equation \ref{precessionrate}) 
and, therefore, will very rapidly become misaligned with each other.

\subsubsection{Misaligned systems from dynamical interactions}

The formation of wide binaries with misaligned discs 
via fragmentation that was described above is the simplest example
of `prompt initial fragmentation' (Pringle 1989) 
where each stellar component and its concomitant disc forms 
from a spatially-distinct region of the collapsing cloud, 
the accretion of large amounts of material on to the binary as a whole 
is avoided (see below), and, thus, the discs need not be initially aligned.

More complicated initial conditions than those described above
can lead to the formation of a small, three-dimensional, cluster 
of stars (e.g.\ Larson 1978; Chapman et al.\ 1993; 
Klessen et al.\ 1998).  The stars in such a group are expected 
to undergo interactions with one another, such as dissipative 
star-disc encounters (Larson 1990; Clarke \& Pringle 1991a, 1991b; 
Clarke \& Pringle 1993; McDonald \& Clarke 1993, 1995).  

Highly-dissipative encounters may lead to the formation of binary 
(or multiple) systems, via capture of the passing object, typically
with misaligned discs.  However, it remains to be determined to 
what extent these dissipative interactions also lead to disc alignment 
through strong tidal interactions (Heller 1993; Hall, Clarke \& Pringle 1996; 
Hall 1997), especially given that in reality
the major infall phase onto the separate protostellar nuclei occurs
contemporaneously with (Bonnell et al.~1997; 
Klessen et al.\ 1998), rather than 
prior to, the dissipative binary formation process (see the next section). 

Another possibility is that the gravitational interaction of a passing 
object, although not leading to capture, may tilt the disc(s) within
a binary system {\em or around a single star}.  If such an interaction
occurred during the main accretion phase, the tilting of the disc
would likely lead to a change in the direction of an emanating jet
which could be mistaken for precession or realignment of a misaligned
disc in a binary system.  Thus, to unambiguously identify a precessing
jet we emphasize that the jet should exhibit several oscillations.

\subsubsection{Subsequent accretion}

Whether a binary is formed directly by fragmentation or more complicated
interactions between several protostars, the above discussion 
ignores the effect of accretion of material after the binary has formed.  

If a protobinary forms 
with discs that are initially misaligned with the orbital plane, 
but subsequently accretes the majority of its mass, the misalignment
will generally be diminished.  For binaries formed directly via
fragmentation, the mass of the protobinary immediately after its 
formation is less for binaries with smaller initial separations 
(Boss 1986).  Therefore, to obtain the same final total mass, closer binaries
must accrete more, relative to their initial mass, than wider binaries
and, hence, close binaries ($\lsim 100$ a.u.) are 
less likely to exhibit misaligned discs than wider binaries.  
Binaries formed via dissipative interactions
in small clusters are also likely to accrete material subsequently
(Bonnell et al.\ 1997; Klessen et al.\ 1998).  Finally, along with these
direct effects of infalling material on misalignment, as mentioned earlier, 
binaries with separations $\lsim 100$ a.u. are also unlikely to 
have strongly misaligned discs because the alignment timescale for 
close binaries ($\lsim 100$ a.u.) is short in comparison to 
typical protostellar accretion timescales.

Conversely, if a binary is formed with discs that lie in the 
orbital plane, there remains the possibility that later infall
of material with a different angular momentum vector to the binary's 
orbit might lead to a degree of misalignment.  
In particular, it is simple for the 
spin of a disc to be affected by late infall of a small 
amount of material once the main
accretion phase is over, and the disc masses have been substantially
reduced. 
On the other hand, the magnitude of the 
misalignment which could occur during the main accretion phase of the stars
in the process of formation, when the jet-like outflows are at their
strongest, and when tidal forces too are at their greatest, has
yet to be determined.  
Further numerical investigation of accretion and fragmentation scenarios 
are required to test this further.

Finally, as with the above mentioned effects of an encounter on the disc
and/or jet of a single star, a change in the angular momentum
vector of material being accreted by a single star could also lead to a
change in the orientation of its disc and, thus,
a wandering of an emanating jet.  Once again, therefore, to unambiguously 
identify a precession, several oscillations must be observed.

\section{Conclusions}

We have investigated the dynamics of a protostellar disc in a binary system 
where the disc is misaligned with the orbital plane of the binary.  
The disc is found to wobble with a period approximately equal to half the
binary's orbital period and to precess with a period of order 20
binary periods.  We also determine the characteristic timescale for
realignment of the disc with the orbital plane due to dissipation.  
If the dissipation is determined by a simple isotropic 
viscosity then we find, in line with previous studies, that the 
alignment timescale is of order the viscous evolution timescale
(of order 100 precession periods).  
However, for typical protostellar disc parameters, 
if the disc tilt exceeds the opening angle of the disc, 
then tidally induced shearing within the disc is transonic.
In general, hydrodynamic instabilities associated with the internally
driven shear result in extra dissipation which is 
expected to drastically reduce the alignment timescale.  For large
disc tilts the alignment timescale is then comparable to the precession
timescale, while for smaller tilt angles $\delta$, the alignment timescale
increases by a factor as $(\sin \delta)^{-1}$.

These general results lead to several observational consequences.
Since the alignment timescale in strongly misaligned discs
is so short, such discs are only likely
to occur in binaries with periods $\gsim 5000$ years (i.e.\ separations
 $\gsim 100$ a.u.).  This expectation is in good agreement with the 
separations of binary systems which are observed to have misaligned jets.  
In addition, because the alignment timescale is
of order the precession period, multiple large wiggles are not expected
to be seen.  At best one might observe a `bending' of the jet as the disc
simultaneously precesses and aligns to the orbital plane.  However, 
such bending could also result from other processes (e.g.\ a single 
star whose disc, and thus jet, orientation is altered by interaction
with a passing object or by the accretion of material with a different
angular momentum vector than that of the original disc).  For this and
other reasons, we emphasize that to unambiguously identify a precessing
jet several oscillations must be observed.  Finally, although multiple
large oscillations are not expected to be seen from systems with 
strongly misaligned discs, the jet direction may be 
forced to wobble by a few degrees on a timescale of half the orbital period, 
with an amplitude that depends on the details of the disc structure.
Such a wobble may have been observed in one of the jets emanating from
the Cep E system and we strongly encourage efforts to determine the binarity
of sources which emit double jets or display evidence of wobbling
jets so that the theoretical expectations can be tested.

For discs which are misaligned by small angles the alignment
timescale is much longer.  Therefore, precession of
jets from such systems may be detectable for systems with orbital periods
of order one hundred years (separations of a few tens of a.u.)
in order to produce multiple wiggles in the $\sim 10^4$ year 
dynamical timescale of a jet.  Wobbling of the jet on a timescale of 
half the orbital period is unlikely to be
large enough to be detected in this case.

Finally, we discuss the implications of the existence of discs that
are misaligned with the orbital plane of a binary for mechanisms for
binary star formation.  
Although most published fragmentation calculations have not
resulted in the formation of discs which are misaligned with the orbital
plane of the binary, it is trivial to produce initial conditions where this 
is the case (e.g.~Bonnell et al.~1992; Bonnell \& Bastien 1992).  
Furthermore, these calculations do not require spatial variation of the
direction of the angular momentum vector in the initial clouds; there is
a single axis of rotation and even solid-body rotation is permitted.  
Binary systems with misaligned discs may also be formed directly 
via dissipative interactions in small clusters of protostars 
formed via `prompt initial fragmentation'.
However, in either case, formation of the binary is likely to be 
contemporaneous with the accretion phase and, thus, strongly misaligned
discs are unlikely for binaries with separations $\lsim 100$ a.u. 
both because of the rapid realignment timescale and because the 
subsequent accretion of a large amount of material by the binary
would tend to align the orbital and disc planes.  Alternatively, binary
systems with misaligned discs could be formed through
the gravitational interaction of a passing object with a previously 
aligned disc, or by the infall of a small amount of material 
with a different angular momentum vector to that of the binary's 
orbit near the end of accretion phase.  Any of these mechanisms could 
explain the misaligned disc which may be present in HK Tau 
(Stapelfeldt et al.~1998; Koresko 1998) provided that the time that has 
elapsed since the misaligned disc was formed is less than 
the current alignment timescale.  Thus, if HK Tau is a binary with
a misaligned disc, this does not necessarily mean that it was
formed this way initially or even that the discs were misaligned during
the main accretion phase when the system would be expected to have
produced jets.  Larger surveys of the alignment of discs in binary 
systems are required for us to draw conclusions on formation mechanisms.

\section*{Acknowledgments}

We thank Jochen Eisl\"offel for his comments on the manuscript.
This work was supported in part by NASA grant
NAG5-4310 and the STScI visitor program.

\end{document}